\documentclass[amsmath,amssymb,aps,pre]{revtex4}
\pdfoutput=1
\usepackage{subfigure}
\usepackage{amsmath}
\usepackage{amssymb}
\usepackage{graphicx}

\newcommand{\mean}[1]{\left \langle #1 \right \rangle}
\newcommand{\parent}[1]{\left( #1 \right)}
\newcommand{\be}{\begin{equation}}
\newcommand{\ee}{\end{equation}}
\newcommand{\bea}{\begin{eqnarray}}
\newcommand{\eea}{\end{eqnarray}}

\begin{document}

\title{\bf Spectral Signature of Nonequilibrium Conditions}

\author{David Andrieux}
\affiliation{
Center for Nonlinear Phenomena and Complex Systems, Universit\'e Libre de Bruxelles, B-1050 Brussels, Belgium. \\
Electronic address: david.andrieux@ulb.ac.be\\
}

\begin{abstract}

The study of stochastic systems has received considerable interest over the years. 
Their dynamics can describe many equilibrium and nonequilibrium fluctuating systems.
At the same time, nonequilibrium constraints interact with the time evolution in various ways.
Here we review the dynamics of stochastic systems from the viewpoint of nonequilibrium thermodynamics. 
We explore the effect of external thermodynamic forces on the possible dynamical regimes and show that the time evolution can become intrinsically different under nonequilibrium conditions. 
For example, nonequilibrium systems with real dynamical components are similar to equilibrium ones when their state space dimension $N < 5$, but this equivalence is lost in higher dimensions. Out of equilibrium systems thus present new dynamical behaviors with respect to their equilibrium counterpart. 
We also study the dynamical modes of generalized, non-stochastic evolution operators such as those arising in counting statistics.
\end{abstract}

\maketitle

\section{Introduction}
\label{Intro}

The concept of thermodynamic equilibrium, first defined as a state where 
no macroscopic changes occur, has gradually evolved over 
the years. In the context of chemical reactions, Wegscheider \cite{W1901} pointed out that the condition of vanishing rate 
does not necessarily coincide with the thermodynamic equilibrium condition. Indeed, stationarity does not rule out the possibility of circular processes, characterized by the presence of matter fluxes. This situation, which 
became known as Wegscheider's paradox, was invoked by Lewis \cite{L25} as 
part of the justification for his general law of entire equilibrium, which 
requires that every elementary process shall have a reverse process and 
that their rates must balance at equilibrium. These assumptions 
were subsequently named microscopic reversibility principle or detailed 
balance conditions. 

It was not, however, until the fifties that the violation of detailed 
balance was quantitatively associated with nonequilibrium properties by 
Klein \cite{K55} in the context of stochastic models. 
Afterwards, Hill \cite{H05} expressed an irreversible entropy production in terms of fluxes and affinities, in analogy with macroscopic thermodynamics. These fluxes and affinities measure the breaking of detailed balance along the cyclic trajectories of the system \cite{H05}. Schnakenberg \cite{S76} and others \cite{LVN84} studied this 
form of the entropy production from a statistical point of view, starting from the Gibbs entropy. 
These works extend a result of Kolmogorov \cite{K36}, who proved that a Markov process is reversible if and only if all 
cyclic trajectories satisfy the detailed balance conditions, to the nonequilibrium realm. 

In parallel, the general research on far-from-equilibrium systems was 
pioneered by Prigogine \cite{GP71, NP77}. In the 
linear regime, equilibrium states are modified by the constraints preventing the system to reach equilibrium, but no new structure appears. 
This situation changes drastically under far-from-equilibrium 
conditions, where coherent space-time behaviors can emerge. 
Spectacular examples are oscillatory Turing patterns in the 
Belouzov-Zhabotinsky reaction or convection rolls in the Rayleigh-B\'enard instability. 
Nonequilibrium conditions can thus be a source of spatio-temporal order. Additional forms of self-organization such as 
the appearance of long-range correlations \cite{KBS02, GM84} and spatial information \cite{NSSN89, AG08p} out of equilibrium have been documented as well.

These progresses uncovered general principles of nonequilibrium self-organization and suggested fundamental differences between equilibrium and nonequilibrium dynamics. Yet, the effect of external thermodynamical driving forces on the time evolution is subtle as, even in their absence, the relaxation towards a steady state is a nonequilibrium dissipative process. 
In this regard, exact mappings between nonequilibrium processes and equilibrium ones exist in different circumstances. 
A first example is found in the chemical network
\begin{eqnarray}
\label{react.1}
\mathrm{A} \rightleftharpoons \mathrm{X} \rightleftharpoons \mathrm{B} \, ,
\label{AXB}
\end{eqnarray}
where the concentrations of species A and B are maintained constant in time. When their chemical potentials are not identical the system is kept out of equilibrium and a flux of matter will flow along the potential gradient. Nevertheless, the time evolution of the system can be mapped onto an equilibrium system of the form $\mathrm{C} \rightleftharpoons \mathrm{X}$ by effectively regrouping the kinetic constants and the two species into a single reaction \cite{AG08c}. More complex, nonlinear reactions networks can be similarly mapped onto equilibrium systems. This is for instance the case of Schl$\ddot{{\rm o}}$gl's model \cite{S72}
\begin{subequations} 
\label{Schlogl}
\begin{eqnarray}
\mathrm{A} +\mathrm{2X} & \rightleftharpoons & \mathrm{3X}  \label{react.1} \\
\mathrm{B} + \mathrm{X} & \rightleftharpoons & \mathrm{C}  \label{react.2} \, ,
\end{eqnarray}
\end{subequations}
which can show multiple steady states and represents a simple model of first-order nonequilibrium transition. 
In both examples several transitions $\rho$ exist between the same physical states (i.e., the transitions $X \rightarrow X \pm 1$ can occur through reaction A or B). These different pathways must be accounted for in the thermodynamic description of the system, but can be merged into a single effective pathway leaving the dynamics invariant. Here the resulting system can be shown to necessarily satisfy the detailed balance conditions (see \cite{AG07} and below) so that the chemical network (\ref{Schlogl}) displays an equilibrium-like dynamics for all concentrations and reaction rates. 

Another class of systems that can be mapped onto an equilibrium motion are driven Brownian motion \cite{A06, A07}. A typical example consists in a Brownian particle trapped in a confining potential $V$, which is experimentally realized by optical tweezers \cite{WSMSE02, AGC}. The system is then driven out of equilibrium by imposing a viscous flow of velocity $u$ and friction coefficient $\alpha$. The resulting drag force moves the Brownian particle away from the potential well and generates a corresponding dissipation. Nonetheless, the Brownian motion can be viewed as an equilibrium random motion in the effective potential $V' = V + \alpha u z$. Here also, the nonequilibrium conditions do not alter the intrinsic dynamics of the system.

Finally, a class of one-dimensional transport models presenting an equilibrium-like dynamics has been recently discovered. Tailleur et al. \cite{TKL07, TKL08} demonstrated that these models can be taken through a non-local transformation into isolated systems satisfying detailed balance. 

These observations bring forth several questions of interest. Is the dynamics intrinsically different under nonequilibrium conditions? If so, what are the qualitative and quantitative differences that may occur and when? 
In this paper we explore these questions in the context of Markovian stochastic systems. 
We first review their spectral properties, using the Jordan-Chevalley decomposition to identify the different dynamical modes. Building on recent results from linear algebra, we study the possible local and global behaviors of these dynamical modes. We show that the presence of thermodynamic driving forces give rise to qualitatively and quantitatively different dynamics.\\
 
\section{Stochastic description}
\label{section.stoch}

In this paper we focus, for the sake of clarity, on Markov chains evolving in discrete time. Similar conclusions can be drawn for continuous-time Markov processes using the uniformization procedure \cite{J53, KT75}.

A discrete-time Markov chain is defined by a probability distribution $p$ over a state space $i \in \{ 1,\ldots, N\}$, and a transition matrix $T_{ij}$ describing the transition probabilities between those states. The transition matrix is non-negative, $T_{ij}\geq 0$, and must satisfy the conditions $\sum_j T_{ij} = 1$ for probability conservation. 

The probability distribution $p$ evolves in discrete time steps according to 
\begin{eqnarray}
p^{(k+1)} = p^{(k)} T \, .
\label{pievol}
\end{eqnarray}
If the Markov chain is irreductible and primitive the probability distribution will evolve towards a unique stationary state satisfying $p^{\rm st} = p^{\rm st} T$ \cite{KT75}. This steady state distribution can be expressed in terms of the minors of the transition matrix \cite{JS92, JQQ04} or in terms of its maximal trees \cite{K1847, S76}.

\section{Dynamical evolution}
\label{section.dynamics}

The iteration scheme 
\begin{eqnarray}
p^{(k)} = p^{(0)}T^k
\label{time.evol}
\end{eqnarray}
solves the time evolution of the probability distribution but a more transparent form can be achieved. 
Using the Jordan canonical decomposition \cite{M01}, a similarity transform $T'=U^{-1}TU$ brings the transition matrix into the form
\begin{eqnarray}
T' = {\rm diag} (J_1, \ldots, J_m) \, ,
\label{J.decomp}
\end{eqnarray}
where $J_i$ is a Jordan block of size $n_i$ and eigenvalue $\lambda_i$:
\begin{eqnarray}
J_i = 
\begin{pmatrix}
\lambda_i & 1       & 0       & \cdots  & 0 \\
0       & \lambda_i & 1       & \cdots  & 0 \\
\vdots  & \vdots  & \ddots& \vdots  & \vdots \\
0       & 0       & 0        & \lambda_i & 1       \\
0       & 0       & 0       & 0       & \lambda_i \\\end{pmatrix} \, .
\end{eqnarray}
Notice that, as the transition matrix $T$ is real, all complex eigenvalues come in conjugate pairs.
The time evolution (\ref{time.evol}) can be written as
\begin{eqnarray}
p^{(k)} &=& p^{(0)}[U {\rm diag} (J_1, \ldots, J_m) U^{-1}]^k \nonumber \\
&=& p^{(0)}U {\rm diag} (J_1^k, \ldots, J_m^k) U^{-1} \, .
\label{time.evol2}
\end{eqnarray}
Noting that the $k^{\rm th}$ power of a Jordan block of size $n$ reads
\begin{eqnarray}
J^k = 
\left(\begin{matrix}
\lambda^k & \binom {k} {1} \lambda^{k-1} & \binom {k} {2} \lambda^{k-2} & \cdots & \binom {k} {n-2} \lambda^{k-n+2} & \binom {k} {n-1} \lambda^{k-n+1} \\
0 & \lambda^k & \binom {k} {1} \lambda^{k-1} & \cdots & \binom {k} {n-3} \lambda^{k-n+3} & \binom {k} {n-2} \lambda^{k-n+2} \\
0 & 0 & \lambda^k & \cdots & \binom {k} {n-4} \lambda^{k-n+4} & \binom {k} {n-3} \lambda^{k-n+3} \\
\vdots & \vdots & \vdots & \ddots & \vdots & \vdots \\
0 & 0 & 0 & \cdots & \lambda^k & \binom {k} {1} \lambda^{k-1} \\
0 & 0 & 0 & \cdots & 0 & \lambda^k \\
\end{matrix}\right) 
\label{Jk}
\end{eqnarray}
we see that the time evolution of the probability distribution can be expressed as
\begin{eqnarray}
p^{(k)} &=& \sum_{i=1}^m \Big[ c^{(i)}_1 \lambda_i^k  + c^{(i)}_2 \binom {k} {1} \lambda_i^{k-1} + \ldots + c^{(i)}_{n_i} \binom {k} {n_i-1} \lambda_i ^{k-n_i +1} \Big] \, .
\label{time.evol3}
\end{eqnarray}
The coefficients $c^{(i)}_l$ are determined by the initial probability distribution $p(0)$ and the generalized eigenvector associated with the decomposition (\ref{J.decomp}).
Accordingly, specifying the set of eigenvalues and Jordan blocks uniquely determines the time evolution of the system.

\begin{center}
\begin{figure}[t] 
\centerline{\scalebox{0.55}{\includegraphics{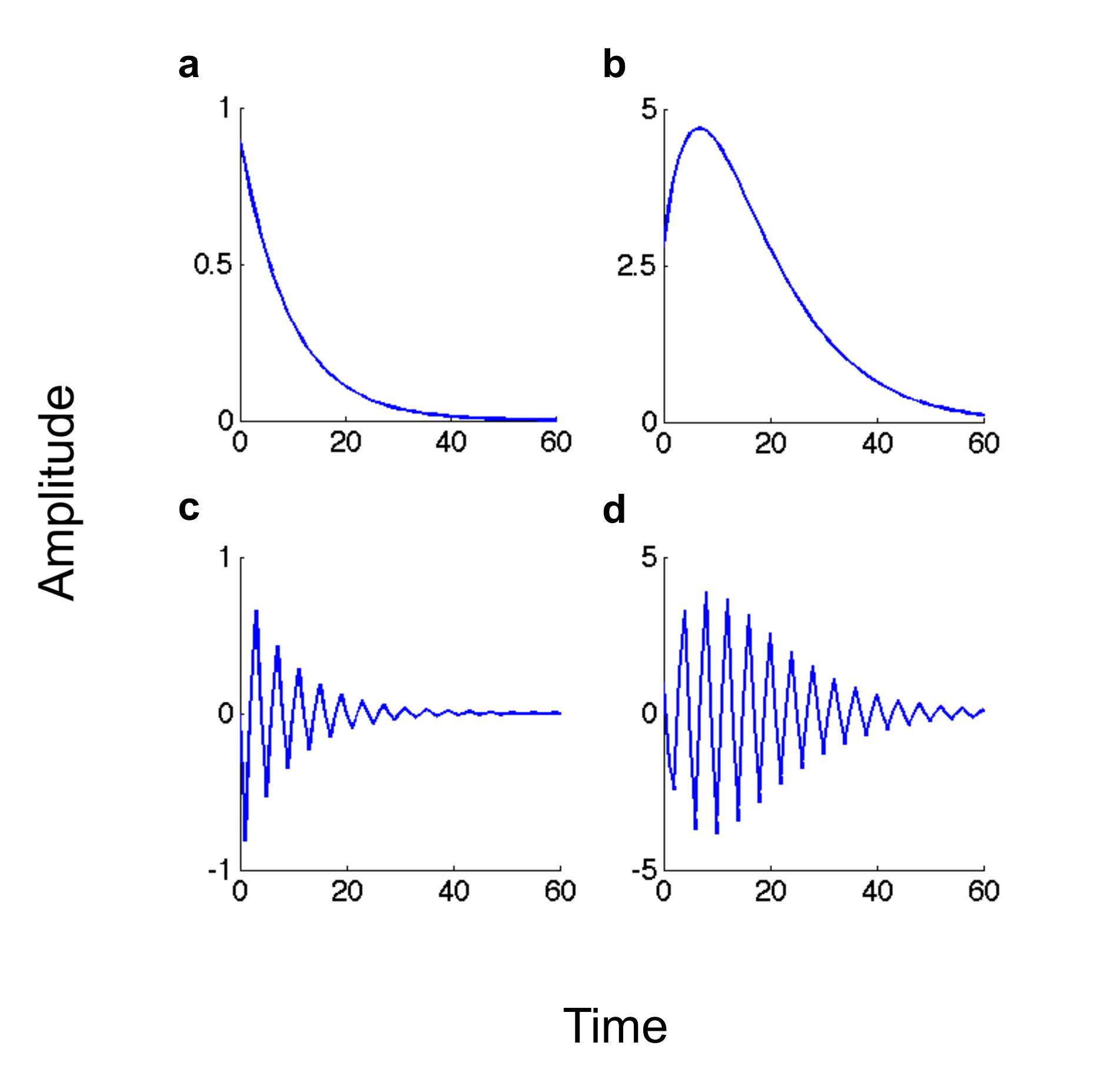}}}
\caption{{\bf Contributions of different dynamical modes to the time evolution.}
 The amplitude of each mode is determined as the real part of the sum of its components over time: $|| J^k || \equiv {\rm Re} \Big[ \sum_{i,j} \parent{J^k}_{ij}$\Big].
In all panels the long-time decay is exponential and associated with an eigenvalue of modulus $|\lambda|=0.9$. 
(a) Exponential mode associated with an eigenvalue $\lambda=|\lambda|$.
(b) Oscillatory mode corresponding to the complex eigenvalue $\lambda =|\lambda| e^{i \pi/2}$. 
(c) Jordan block of eigenvalue $\lambda=|\lambda|$ and size $2$.
(d) Complex Jordan block of eigenvalue $\lambda = |\lambda| e^{i \pi/2}$ and size $2$.} 
\label{fig1} 
\end{figure}
\end{center} 

Each of these contributions has a different impact on the time evolution.
We can distinguish four classes of dynamical behaviors as follows: an eigenvalue is either real or complex, and occurs either in a Jordan block of size $1$ or larger. Their respective influence on the dynamical evolution is depicted in Figure \ref{fig1}.
In addition to an exponential decay, we see that the presence of a complex eigenvalue (and its conjugate) leads to an oscillatory behavior. In parallel, the presence of a Jordan block of size greater than one (here $n=2$) leads to a large transient component, which can be coupled to an oscillatory component or not. These behaviors arise from the terms $\propto k^l \lambda^{k}$ ($l=0,\ldots,n-1$) in equation (\ref{Jk}) or (\ref{time.evol3}). Such transients thus become longer with the size of the Jordan block. Similar results hold in continuous-time as well \cite{Note}.

The eigenvalues obey the Perron-Frobenius theorem \cite{M01}:\\

If $A$ is a $N \times N$ non-negative, primitive, and irreducible matrix, then

          1) there is a real eigenvalue $\lambda_1$ of $A$ such that any other eigenvalue $\lambda$ satisfies $|\lambda| < \lambda_1$. 

          2) there is a left (respectively right) eigenvector associated with $\lambda_1$ having positive entries.

    3) that eigenvalue is a simple root of the characteristic equation of A.\\

Probability conservation requires that the rows of a transition matrix sum to unity, which in turn implies that the Perron root $\lambda_1=1$. The other eigenvalues can be ordered as $\lambda_1 > |\lambda_2| \geq \cdots \geq |\lambda_r|$. The eigenvalue $\lambda_1=1$ corresponds to the stationary state while the second largest eigenvalue $\lambda_2$ characterizes the relaxation towards the steady state. 
Importantly, Karpelevic characterized the complex plane region $\Theta_N$ realizable by the eigenvalues of $N \times N$ stochastic matrices \cite{K51,M88}. This result does not, however, describe the set of possible {\it combinations} of eigenvalues. 

\section{Thermodynamic description}

Based on the concept of entropy production, a thermodynamic description has been associated with Markov processes \cite{K55, H05, S76, NP77}. A system is out of equilibrium at time $k$ if the irreversible entropy production
\begin{eqnarray}
\Delta_{\rm i} S^{(k)} = \frac{1}{2}\sum_{i,j} \parent{ p^{(k)}_i T_{ij} - p^{(k)}_j T_{ji}  } \ln \parent{ \frac{p^{(k)}_i T_{ij} }{p^{(k)}_j T_{ji}} } \geq 0
\label{dis}
\end{eqnarray}
is positive. This is typically the case when the system relaxes towards its steady state. 
The relaxation process can be driven by internal (due to a non-stationary initial probability distribution) and external (e.g., chemical potential gradients between reservoirs) thermodynamic forces. 
At the steady state the entropy production may vanish, in which case the system has reached a state of thermodynamic equilibrium, or present a positive value, in which case the system is maintained out of equilibrium by the presence of external thermodynamic forces acting on the system.

As seen from equation (\ref{dis}), a stationary state is an equilibrium state when the conditions of detailed balance
\begin{eqnarray}
p^{\rm eq}_i T_{ij} = p^{\rm eq}_j T_{ji}  
\label{DB}
\end{eqnarray}
are satisfied for all the possible forward and backward transitions. That is, every elementary process has a reverse process, and their rates balance at equilibrium \cite{W1901, L25}.
In particular, this implies that no probability flux is present at equilibrium. 
Note that, in the present form, the knowledge of the equilibrium distribution is required to verify whether a system satisfies detailed balance at the steady state.

An equivalent condition, known 
as Kolmogorov's criterion \cite{K36}, can be expressed in terms of the sole transition probabilities. The equilibrium conditions (\ref{DB}) are equivalent to:\\

(1) $T_{ij} > 0$ implies $T_{ji}  > 0$.

(2) $T_{i_1 i_2} \ldots T_{i_n i_1} = T_{i_1 i_n} \ldots T_{i_2 i_1 }$ for any finite sequence $(i_1, i_2, \ldots, i_n)$.\\

These conditions are necessary and sufficient in order for the steady state of the system to be at equilibrium, i.e. a Markov process is reversible if and only if all cyclic trajectories satisfy the detailed balance conditions. A direct consequence of this observation is that systems without cyclic paths (other than the $(i_1, i_2)$ pairs) are always at equilibrium, regardless of the transition probabilities. For example, the system $1 \rightleftharpoons 2 \rightleftharpoons 3$ satisfies the detailed balance conditions for all (positive) transition probabilities. 

The expressions for the entropy production (\ref{dis}) and the detailed balance conditions (\ref{DB}) assume that there is a unique transition pathway between states. When several transition pathways $\rho$ exist between states, as in the chemical networks (\ref{AXB}) or (\ref{Schlogl}), it is necessary to include them in the thermodynamic formulation. 
The expressions (\ref{DB}) and (\ref{dis}) must be adjusted as follows. The detailed balance conditions (or Kolmogorov's conditions) must hold for all individual pathways and transitions: $p^{\rm eq}_i T^{(\rho)}_{ij} = p^{\rm eq}_j T^{(\rho)}_{ji} $. Similarly, the sum over all transitions in the entropy production must now include an additional sum over all possible transition pathways.   

Kolmogorov's conditions are equivalent to the symmetrizability of the transition matrix (i.e., it is similar to a symmetric matrix).
In this case the transition matrix is diagonalizable and all its eigenvalues are real. We will use this important observation in our subsequent analysis. 
The mapping to a symmetric matrix is readily achieved if the stationary distribution is known. Indeed, the similarity operator 
\begin{eqnarray}
U={\rm diag} \parent{ \sqrt{p_1}, \ldots, \sqrt{p_n} }
\end{eqnarray}
brings $T$ into the form
\begin{eqnarray}
T'_{ij} = (U T U^{-1})_{ij} = \sqrt{\frac{p_i}{p_j}} T_{ij} \, .
\end{eqnarray}
The latter is symmetric,  
\begin{eqnarray}
T'_{ij} = \sqrt{T_{ij}T_{ji}} = T'_{ji} \, , 
\end{eqnarray}
when $p=p^{\rm eq}$ satisfies the detailed balance conditions (\ref{DB}). 
Alternatively, a similarity transformation independent of the equilibrium distribution $p^{\rm eq}$ has been derived that brings the system into a symmetric form using only the transition probabilities and the topology of the chain \cite{AG08}.

When the detailed balance conditions are not satisfied,
\begin{eqnarray}
p^{\rm st}_i T_{ij} \neq p^{\rm st}_j T_{ji}   \, ,
\end{eqnarray}
the steady state defines a nonequilibrium stationary state. 
In this case probability fluxes will flow through the system and generate thermodynamic forces or affinities \cite{S76}. 
%
In this way, Kolmogorov's relations, which do not depend on the probability distribution, provide an intrinsic measure of the thermodynamic conditions.
In the following we thus refer to systems satisfying Kolmogorov's conditions as presenting an equilibrium-based dynamics. 
We emphasize that, even in this case, the relaxation process towards the equilibrium distribution is a genuine nonequilibrium process characterized by a positive entropy production.

\section{Equilibrium versus nonequilibrium spectrum}

The characterization of a Markov process as presenting an equilibrium- or nonequilibrium-based dynamics rests on the absence or presence of external thermodynamic forces. The thermodynamic conditions can be revealed by Kolmogorov's relations, which are based on the value of the transition probabilities along the cyclic paths of the chain. Similarly, the dynamical evolution results from a complex interplay between the topology of the system and its transition probabilities or effective connectivity. 
In this regard, the effects of thermodynamic driving forces on the dynamics itself are difficult to predict. Here the question we want to investigate is the following: Does the presence of external thermodynamical forces change the nature of the dynamics, and in what ways? 
To address this question we identify the possible dynamical regimes with or without external thermodynamical driving forces. We show that the nonequilibrium conditions determine different classes of dynamical behaviors with respect to an equilibrium-based dynamics. 

A first observation is that we can restrict our analysis to the case where only one type of transition exists between two states. Indeed, although the presence of several transitions affects the thermodynamic properties of the system (e.g., the entropy production), it is possible to merge all these different pathways onto a unique, effective transition without affecting the dynamics itself. In some cases the resulting system can be at equilibrium if it satisfies the detailed balance conditions for the total transition probabilities $T_{ij} \equiv \sum_\rho T^{(\rho)}_{ij}$. When the resulting system is at equilibrium for all values of the transition probabilities (e.g., because it does not present cyclic trajectories), we have a robust class of mapping from nonequilibrium systems to equilibrium ones. This is for example the case for the chemical networks (\ref{AXB}) and (\ref{Schlogl}) discussed in the Introduction.

We can now distinguish different situations, based on the Jordan-Chevalley decomposition of the transition matrix. 
As discussed in Section \ref{section.dynamics}, this scheme provides us with the set of eigenvalues and Jordan blocks characteristics of the system and its dynamics.

\begin{center}
\begin{figure}[t] 
\centerline{\scalebox{0.55}{\includegraphics{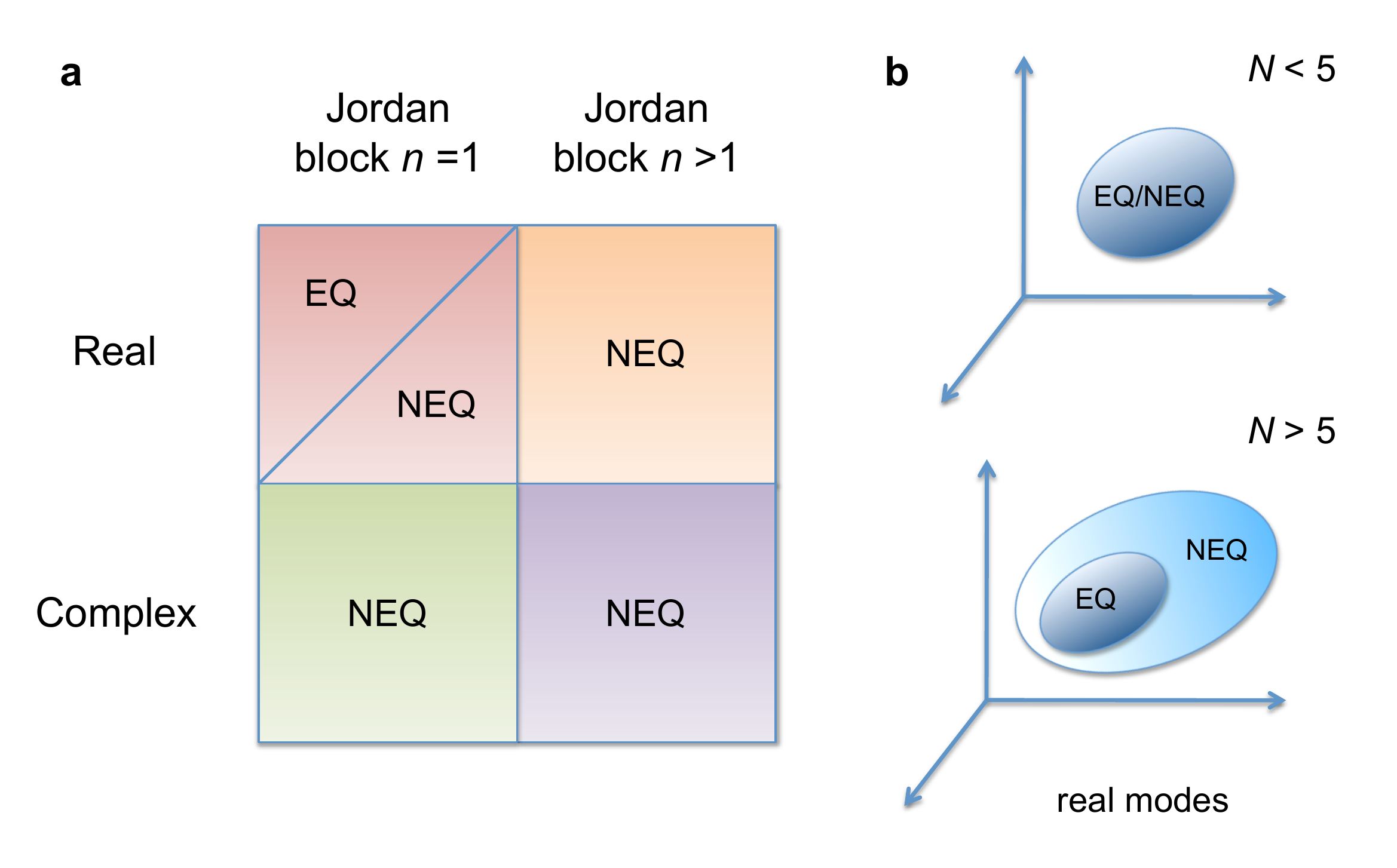}}}
\caption{{\bf Classification of dynamical modes in terms of thermodynamic conditions.}
(a) The presence of a complex eigenvalue or a Jordan block of size $>1$ is characteristic of a nonequilibrium process. When all eigenmodes are real, the thermodynamic conditions are uncertain. 
(b) Subspace of real dynamical modes. When the state space dimension $N<5$, the combinations of real modes accessible under equilibrium or nonequilibrium conditions are identical. When the state space dimension $N>5$, some combinations of real modes only occur under nonequilibrium conditions.} 
\label{fig2} 
\end{figure} 
\end{center}

We first consider equilibrium-based dynamics satisfying Kolmogorov's conditions. In this case the transition matrix is symmetrizable so that its eigenvalues are real and the eigenvectors form a complete basis. In particular, this implies that neither complex eigenvalues nor Jordan blocks can occur when no thermodynamic forces are present. Oscillating dynamical modes and 'slow' transients arising from Jordan blocks can only occur in presence of external thermodynamic forces (see Figure \ref{fig2}a), in contrast with an equilibrium-based dynamics. Another direct consequence is that the equilibrium power spectrum is Lorentz-typed \cite{QQZ03}; in particular no stohastic resonance can occur at equilibrium. This observation already reveals that equilibrium and nonequilibrium systems can exhibit fundamentally different dynamics.

The remaining case to consider occurs when all eigenvalues are real and when no Jordan blocks are present. The system is then entirely characterized by the set $(\lambda_1, \lambda_2, \ldots, \lambda_r)$ of eigenvalues.  In this situation it is more difficult to distinguish equilibrium- from nonequilibrium-based dynamics. 
We thus investigate the relationships between the set of eigenvalues $(\lambda_1, \ldots, \lambda_r)$ of a stochastic matrix and its thermodynamic properties. As mentioned earlier, Karpelevic's theorem \cite{K51,M88} restricts the domain of possible eigenvalues, but it does not describe the possible combinations of eigenvalues of stochastic matrices. This difficult  task is known as the inverse eigenvalue problem \cite{C98}: given a set of eigenvalues and a space of operators, can we find a system that realizes this particular set of eigenvalues? 

In our case, the class of operators of interest corresponds to the space of stochastic matrices with real eigenvalues. However, it is easier to study the inverse eigenvalue problem in the space of non-negative operators, referred to as the real non-negative inverse eigenvalue problem (RNIEP). The following mapping then relates irreductible non-negative matrices to stochastic ones. 
Consider a non-negative matrix $A$ and denote its Perron root by $\lambda_1$ and its (entrywise positive) right Perron vector by $x$. Then if $D = {\rm diag}(x_1, \ldots, x_N)$ we have that \cite{DD45}
\begin{eqnarray}
T = \frac{1}{\lambda_1} D^{-1} A D
\label{mapT}
\end{eqnarray}
is a stochastic matrix. 
For our purpose, an additional observation must be made: if $A$ satisfies Kolmogorov's conditions then its derived stochastic matrix $T$ also does. 
This is shown as follows. The transition matrix has elements $T_{ij}=\lambda^{-1}_1 D^{-1}_{ik} A_{kl}  D_{lj}=  \lambda^{-1}_1 (1/x_i) \delta_{ik} A_{kl}  \delta_{lj} x_j=  \lambda^{-1}_1 (x_j/x_i) A_{ij}$ so that
\begin{eqnarray}
T_{i_1 i_2} \ldots T_{i_n i_1} = \lambda^{-n}_1 A_{i_1 i_2} \ldots A_{i_n i_1}
\end{eqnarray}
for any cycle $(i_1, i_2, \ldots, i_n)$, from which we deduce that $T_{i_1 i_2} \ldots T_{i_n i_1} = T_{i_1 i_n} \ldots T_{i_2 i_1 }$.
This shows that the construction (\ref{mapT}) preserves the distinction between equilibrium- and nonequilibrium-based dynamics. Consequently, we can study the eigenvalues of non-negative matrices and transfer the results to stochastic matrices.

We can now restate the problem as follows: Can we find a non-negative operator realizing a given set of real eigenvalues? More importantly, are there some combinations of eigenvalues that can be realized by a non-negative matrix but not by a symmetric one? Partial answers to these questions have started to emerge recently \cite{C98, ELN04}. Here we use recent developements in linear algebra to show that nonequilibrium conditions allow for new patterns of real dynamical modes.

Partial results for the inverse eigenvalue problem on stochastic matrices were first obtained by Soule \cite{S83} (see also \cite{CD91}). Later on, the RNIEP problem was solved by Loewy and London when $N < 5$ \cite{LL78}. More recently, the problem over the set of symmetrizable matrices was solved by Wumen \cite{W97}, also for $N < 5$. Notably, they found the same subset in the space of eigenvalues, implying that these two classes of problems are equivalent for $N < 5$. This means that systems with low-dimensional state spaces determine the same class of real dynamics, regardless of the thermodynamic conditions. 

For higher dimensions, Johnson and coworkers eventually demonstrated that the two classes of problems are not equivalent, at least in high dimensions \cite{JLL96}. A final step was accomplished when Egleston derived a class of eigenvalues combinations that cannot be realized by any $5 \times 5$ symmetrizable matrix \cite{E01}.
For example, the set of eigenvalues $\parent{1, \frac{71}{97}, -\frac{44}{97}, -\frac{54}{97}, -\frac{70}{97} }$ is realizable but not by a symmetrizable matrix.
We can thus conclude that, for $N \geq 5$, the nonequilibrium conditions extend the range of possible dynamical patterns (Figure \ref{fig2}b). This completes the analysis of equilibrium- versus nonequilibrium-based dynamics.

\section{Time evolution of thermodynamic fluctuations}

The dissipation rate and the thermodynamic currents play an important role in nonequilibrium statistical thermodynamics \cite{S76, NP77}. The dissipation is related to the irreversible entropy production and the efficiency of free energy conversion into useful work. The thermodynamic currents describe the fluxes of matter or energy flowing through the system. Their response and fluctuation properties are therefore of fundamental interest, especially, for the exploration of nanoscale systems.

The dynamical properties of such thermodynamic quantities have been recently investigated (see \cite{SPWS08} for a review). They are described by the so-called generating functions, which evolve in time according to generalized transition operators. Although these operators are not row-stochastic (i.e., their rows don't sum to unity), they are non-negative so that a similar analysis can be performed. 
The natural question is thus whether these thermodynamic quantities present a fundamentally different dynamical character than the original state space dynamics. Interestingly, the answer turns out to depend on the particular quantity under study.

\begin{center}
\begin{figure}[t] 
\centerline{\scalebox{0.55}{\includegraphics{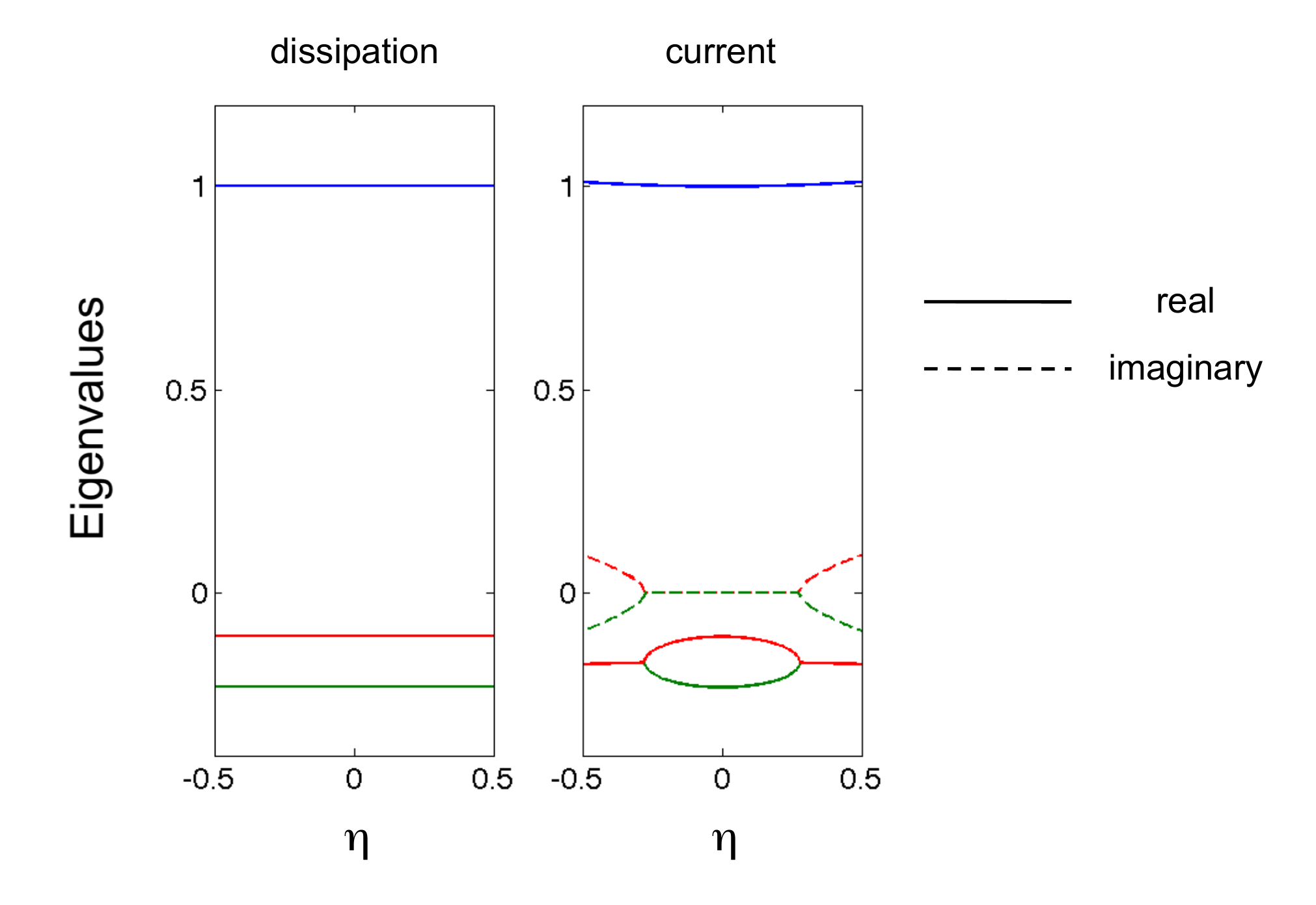}}}
\caption{{\bf Dynamical modes of the dissipation rate and thermodynamic currents of the cyclic system $1 \rightleftharpoons 2 \rightleftharpoons 3 \rightleftharpoons 1$ at equilibrium.} The transition probabilities take the values $T_{12}=T_{13}=0.42, T_{21}=T_{23}=0.35, T_{31}=T_{32}=0.4$ and $T_{ii}=1-\sum_{j\neq i} T_{ij}$, which satisfy Kolmogorov's conditions. The current is measured as $\epsilon_{13}=-\epsilon_{31}=1$ and zero otherwise.
Left panel: Eigenvalues of the dissipation rate fluctuations operator (\ref{T.dis}). They are real for all values of the parameter $\eta$. Right panel: Eigenvalues of the current fluctuations operator (\ref{T.J}). They present (conjugate) complex components for $|\eta| > \eta_c$, indicating an oscillating behavior in the time evolution.} 
\label{fig3} 
\end{figure} 
\end{center}

We first consider the entropy production fluctuations of Markov stochastic processes. 
They are described by a generating function, obtained as the Fourier transform $\mean{\exp(-\eta S(k))}$ of the probability distribution of observing a given value of the total dissipation $S(k)\equiv \sum_{l=1}^k \log(T_{i_l i_{l+1}}/T_{i_{l+1}i_l})$ along a trajectory of length $k$.
The operator describing the time evolution of the generating function reads \cite{K98, LS99}
\begin{eqnarray}
L_{ij} (\eta)=  T^{1-\eta}_{ij} T^{\eta}_{ji} \, .
\label{T.dis}
\end{eqnarray}
We recover the evolution operator of the probability distribution when $\eta=0$.
It is readily verified that, when the system satisfies Kolmogorov's conditions, this generalized operator describing the time evolution of the entropy production fluctuations also satisfies these conditions. Indeed, for any cycle $(i_1, i_2, \ldots, i_n)$ we have
\begin{eqnarray}
L_{i_1 i_2} \ldots L_{i_n i_1} (\eta)
&=&  T_{i_1 i_2} \ldots T_{i_n i_1}  \parent{\frac{ T_{i_1 i_n} T_{i_n i_{n-1}} \ldots T_{i_2 i_1} }{ T_{i_1 i_2} \ldots T_{i_n i_1} }  }^{\eta} \nonumber \\
&=&  T_{i_1 i_2} \ldots T_{i_n i_1}
\end{eqnarray}
where we used Kolmogorov's condition $T_{i_1 i_2} \ldots T_{i_n i_1} = T_{i_1 i_n} T_{i_n i_{n-1}} \ldots T_{i_2 i_1}$. 
By virtue of the latter equality we also have
\begin{eqnarray}
L_{i_1 i_2} \ldots L_{i_n i_1} (\eta) =  L_{i_1 i_n} L_{i_n i_{n-1}} \ldots L_{i_2 i_1} (\eta) 
\end{eqnarray}
for all values of $\eta$.
Consequently the operator (\ref{T.dis}) is symmetrizable and its eigenvalues are real with no Jordan blocks of size $n>1$ \cite{Note2}.
Therefore, the entropy production and its equilibrium fluctuations always present a time evolution quantitatively similar to the probability distribution, i.e. characterized by exponential decay modes belonging to the same set as those of the probability distribution itself.

The situation differs considerably for the thermodynamic currents.
For our purpose it is sufficient to consider the case where only one thermodynamic current is present. In this case the total current along a trajectory is obtained as $G(k)\equiv \sum_{l=1}^k \epsilon_{i_l i_{l+1}}$, where $\epsilon_{ij}=-\epsilon_{ji}$ takes the value $\pm 1$ if the transition $i \rightarrow j$ generates a positive (negative) current, and zero otherwise.
The operator describing the time evolution of the generating function $\mean{\exp(-\eta G(k))}$ reads \cite{AG04, AG07}
\begin{eqnarray}
L_{ij} (\eta)=  T_{ij} e^{-\epsilon_{ij} \eta} \, .
\label{T.J}
\end{eqnarray}
Even for systems at equilibrium this evolution operator does not, however, satisfies Kolmogorov conditions. For example, a cycle associated with a single current pulse would yield
\begin{eqnarray}
L_{i_1i_2} \ldots L_{i_n i_1} (\eta)
&=&  T_{i_1i_2} \ldots T_{i_n i_1} e^{- \eta} \nonumber \\
&=&  T_{i_1i_n} T_{i_n i_{n-1}} \ldots T_{i_2 i_1} e^{- \eta} \nonumber \\
&\neq&  T_{i_1i_n} T_{i_n i_{n-1}} \ldots T_{i_2 i_1} e^{+\eta} =  L_{i_1i_n} L_{i_n i_{n-1}} \ldots L_{i_2 i_1} (\eta)
\end{eqnarray}
for $\eta \neq 0$.
The thermodynamic currents may thus display more complex behaviors (slow transients, oscillations) than those observed in the state space time evolution, even for equilibrium systems. This is illustrated in Figure \ref{fig3} where we observe, on the simple system $1 \rightleftharpoons 2 \rightleftharpoons 3 \rightleftharpoons 1$, that the current operator can present complex modes.
Thus, the present approach also serves as a basis to study and classify the dynamical properties of further thermodynamical quantities of interest.\\

\section{Conclusions}
\label{conclusion}

The dynamics of stochastic systems emerges from a complex interplay between their topology and their effective connectivity (as measured by the transition probabilities). On the other hand, nonequilibrium conditions generate dissipation that in turn affects the time evolution. Interestingly, the presence of external thermodynamic forces is revealed at the level of the effective connectivity, as described by Kolmogorov's conditions. 

Intrinsic differences between equilibrium-based and nonequilibrium-based dynamics can be uncovered. 
These differences may take different, possibly coexisting, forms. 
First, complex dynamical modes can only appear out of equilibrium. They generate oscillatory components in the time evolution (Figures \ref{fig1}c, d). 
Second, the presence of Jordan blocks only occurs under nonequilibrium conditions. These contributions give rise to slow dynamical components in the time evolution (Figures \ref{fig1}b, d). 


When none of those phenomena are present, the time evolution displays a uniform relaxation towards the steady state. 
The combination of then real eigenvalues determines the dynamics of the system. Notably, when the state space is low-dimensional ($N < 5$) the possible combinations of eigenvalues are strictly identical for equilibrium and nonequilibrium systems. In this sense they are indistinguishable and the nonequilibrium conditions do not increase the range of possible behaviors. In contrast, when $N \geq 5$ the diversity of possible patterns is enlarged (Figure \ref{fig2}b) and additional dynamical behaviors become possible. 

We have also considered the fluctuations of thermodynamical quantities such as the dissipation rate and the thermodynamic currents. Their time evolution can be described by non-negative operators similar to a transition matrix, but with the important difference that their largest eigenvalue is diffeerent from one. 
The resulting equilibrium dynamical behaviors can be similar to the time evolution of the state space probability distribution, as for the dissipation rate, or intrinsically different, as for the thermodynamic currents (Figure \ref{fig3}). In particular, the thermodynamic fluxes can present a nonequilibrium-like behavior even when the underlying system is at equilibrium.


These differences may, at first glance, appear surpising. Indeed, for a given system, the set of admissible trajectories is identical under equilibrium and nonequilibrium alike. The only difference comes from the statistical weight given to these trajectories. Yet the dynamical modes may become qualitatively different under nonequilibrium conditions. The detailed connection between the dynamical modes, which are global quantities, and the trajectories remains to be explored. 

Further interesting questions can be envisaged based on the present results. 
For instance, experimental observations often offer limited information on the system. The present approach may be used to discriminate between equilibrium and nonequilibrium dynamics on the basis of data obtained from a subset of the total phase space (see also \cite{ASS10}).
Indeed, observing a subset of the full state space can be sufficient to identify several of the dynamical modes involved and thus distinguish between a global equilibrium or nonequilibrium state. 
Determining the thermodynamic conditions is especially relevant in the study of biochemical networks, enzyme kinetics, or molecular motors.\\

{\bf Acknowledgments.} 
This work is supported by the F.R.S-FNRS Belgium. 
Part of this work was accomplished at the Department of Neurobiology and Kavli Institute for Neuroscience, Yale University School of Medicine, New Haven, CT, USA.


\begin{thebibliography}{99}
\bibitem{Note} In continuous-time, the time evolution operator is expressed as the exponential $\exp(tT)$ of the transition matrix. A Jordan block of size $n$ thus gives rise to contributions of the form $t^{k} e^{-\lambda t}$ to the time evolution, with $k=0,\ldots, n-1$. The factors $t^k$ generate slow transients in the time evolution before being dominated by the exponential decay.
\end{thebibliography}
\end{document}